\documentclass[11pt,a4paper,american]{extarticle}
\usepackage[T1]{fontenc}
\usepackage[latin1]{inputenc}
\usepackage{amsmath}
\usepackage{amssymb}
\usepackage{setspace}
\usepackage{esint}
\makeatletter

\providecommand{\tabularnewline}{\\}

\usepackage{babel}
\usepackage{bbm}
\usepackage{hyperref}
\usepackage{authblk}
\date{}
\allowdisplaybreaks
\numberwithin{equation}{section}

\author[1]{Thales Azevedo\thanks{thales.azevedo@physics.uu.se}}
\author[2]{Renann Lipinski Jusinskas\thanks{renannlj@fzu.cz}}
\affil[1]{Department of Physics and Astronomy, Uppsala University \authorcr Box 516, 751 20,  Uppsala - Sweden\authorcr \ }
\affil[2]{Institute of Physics AS CR \authorcr  Na Slovance 2, 182 21, Prague - Czech Republic}

\makeatother

\usepackage{babel}
\begin{document}

\title{Background constraints in the infinite tension limit of the heterotic string}
\maketitle
\begin{abstract}
In this work we investigate the classical constraints imposed on the
supergravity and super Yang-Mills backgrounds in the $\alpha'\to0$
limit of the heterotic string using the pure spinor formalism. Guided
by the recently observed sectorization of the model, we show that
all the ten-dimensional constraints are elegantly obtained from the
single condition of nilpotency of the BRST charge.

\tableofcontents{}
\end{abstract}

\section{Introduction\label{sec:Introduction}}

\

About three years ago, Cachazo, He and Yuan (CHY) proposed a compact
formula for computing tree-level amplitudes in both Yang-Mills and
gravity theories \cite{Cachazo:2013hca}. There was an increasing
interest then to find a string origin of those results given their
known connection to string amplitudes at the low-energy limit.

Soon after that work, Mason and Skinner introduced the so-called ambitwistor
string \cite{Mason:2013sva}, which could be viewed as an $\alpha'\to0$
limit of the usual string and provided a clear derivation of the CHY
formulae for $D=10$ Yang-Mills and NS-NS supergravity.

Taking advantage of the pure spinor formalism's manifest supersymmetry,
Berkovits proposed its ambitwistor version in \cite{Berkovits:2013xba},
which was explicitly shown in \cite{Gomez:2013wza} to provide the
supersymmetric version of the CHY amplitudes.

When extended to curved backgrounds, one would expect that consistency
of the ambitwistor string should put  the target
space fields  on-shell. In \cite{Adamo:2014wea}, Adamo \emph{et al} demonstrated
that the nonlinear equations of motion of the NS-NS background arise
as anomalies of the worldsheet supersymmetry algebra. In the pure
spinor case, Chandia and Vallilo investigated the type II background
\cite{Chandia-Vallilo} and realized that Berkovits' original proposal
for the infinite tension string was incomplete and had to be modified
in order to obtain the usual background constraints coming from
the pure spinor formalism. By performing a semi-classical analysis,
they were able to reproduce the known results of \cite{Berkovits:2001ue}
with the introduction of the extra condition of BRST-closedness of
$\mathcal{H}$, a generalized particle-like Hamiltonian.

The ideas in \cite{Chandia-Vallilo} were further explored by one
of the authors in \cite{Jusinskas:2016qjd} and it was shown that
the new model, although still chiral, could be interpreted in terms
of two sectors resembling the usual left and right-movers of the superstring.
This construction was also extended to the heterotic case, providing
a sensible description of the massless heterotic spectrum in this
$\alpha'\to0$ limit. This was achieved by incorporating the observed
sectorization in the heterotic BRST charge, which was then redefined
to be

\begin{equation}
Q=\oint\{\lambda^{\alpha}d_{\alpha}+\bar{c}T_{+}-\bar{b}\bar{c}\partial\bar{c}\},\label{eq:BRSTintro}
\end{equation}
where $\lambda^{\alpha}$ is the pure spinor ghost, $d_{\alpha}$
is the improved worldsheet realization of the superderivative introduced
in \cite{Chandia-Vallilo}, ($\bar{b},\bar{c}$) are the
reparametrization ghosts and $T_{+}$ accounts for one of the sectorized
energy-momentum-like tensors, which are defined in terms of $\mathcal{H}$
and the full energy-momentum tensor $T$ as
\begin{equation}
T_{\pm}\equiv\frac{1}{2}(T\pm\mathcal{H}).
\end{equation}

As we show in the present work, the problem of finding the constraints
on the heterotic background is somewhat more natural than in type
II, in that $\mathcal{H}$ enters the BRST charge $Q$ itself, \emph{cf}.
\eqref{eq:BRSTintro}, and the background constraints all come from
the sole requirement that $Q$ be nilpotent. In a general heterotic
background, the action for the sectorized model and the generalized
particle-like Hamiltonian will be cast as
\begin{eqnarray}
S & = & \frac{1}{2\pi}\int d^{2}z\{\mathcal{P}_{a}\bar{\Pi}^{a}+d_{\alpha}\bar{\Pi}^{\alpha}-\Pi^{A}\bar{\Pi}^{B}B_{BA}+\bar{\Pi}^{A}A_{A}^{I}J_{I}+w_{\alpha}\bar{\nabla}\lambda^{\alpha}+\bar{b}\bar{\partial}\bar{c}\}+S_{C},\label{eq:actionintro}\\
\mathcal{H} & = & -\frac{1}{2}\mathcal{P}_{a}\mathcal{P}^{a}-\frac{1}{2}\Pi^{a}\Pi_{a}+d_{\alpha}\Pi^{\alpha}+w_{\alpha}\nabla\lambda^{\alpha}-\bar{b}\partial\bar{c}-\partial(\bar{b}\bar{c})+T_{C}\nonumber \\
 &  & -\Pi^{A}A_{A}^{I}J_{I}-d_{\alpha}W^{\alpha I}J_{I}-\lambda^{\alpha}w_{\beta}U_{\alpha}^{\hphantom{\alpha}\beta I}J_{I}.
\end{eqnarray}
The vielbein appears through $\Pi^{A}=\partial Z^{M}E_{M}^{\hphantom{M}A}$,
mapping the curved superspace coordinates $Z^{M}$, to the generalized
superspace invariants with flat (super) indices $A$. The Lorentz
connection $\Omega_{AB}^{\hphantom{AB}C}$, enters the covariant derivative
$\nabla$. The super Kalb-Ramond field is denoted by $B_{AB}$, while
$A_{A}^{I}$, $W^{\alpha I}$ and $U_{\alpha}^{\hphantom{\alpha}\beta I}$
represent the super Yang-Mills background. All the worldsheet fields
above will be detailedly introduced in section \ref{sec:heterotic}.

By performing a classical analysis and computing the generalized Poisson
brackets associated to $S$, we will show that classical nilpotency
of the BRST charge \eqref{eq:BRSTintro} imposes some constraints
on the torsion $T_{AB}^{\hphantom{AB}C}$, the 3-form field strength
$H_{ABC}$, the curvature tensor $R_{ABC}^{\hphantom{ABC}D}$, and
the super Yang-Mills field strength $F_{AB}^{I}$, given by\begin{subequations}\label{eq:hetconstraints}
\begin{equation}
\lambda^{\alpha}\lambda^{\beta}T_{\alpha\beta}^{\hphantom{\alpha\beta}A}=\lambda^{\alpha}\lambda^{\beta}H_{A\alpha\beta}=\lambda^{\alpha}\lambda^{\beta}\lambda^{\gamma}R_{\alpha\beta\gamma}^{\hphantom{\alpha\beta\gamma}\delta}=\lambda^{\alpha}\lambda^{\beta}F_{\alpha\beta}^{I}=0,
\end{equation}
in addition to the so-called holomorphicity constraints\footnote{This name can be misleading here, as the infinite tension limit is
described by the chiral action \eqref{eq:actionintro} and holomorphicity
of the BRST current is trivial.}
\begin{equation}
T_{\alpha a}^{\hphantom{\alpha a}\beta}=T_{\alpha(ab)}=T_{\alpha\beta b}-H_{\alpha\beta b}=H_{ab\alpha}=\lambda^{\alpha}\lambda^{\beta}R_{\alpha a\beta}^{\hphantom{\alpha a\beta}\gamma}=0,
\end{equation}
and
\begin{eqnarray}
F_{\alpha a}^{I} & = & T_{\alpha\beta a}W^{\beta I},\\
\nabla_{\alpha}W^{\beta I}-T_{\alpha\gamma}^{\hphantom{\alpha\gamma}\beta}W^{\gamma I} & = & U_{\alpha}^{\hphantom{\alpha}\beta I},\\
F_{\alpha\beta}^{I} & = & \frac{1}{2}W^{\gamma I}H_{\alpha\beta\gamma},\\
\lambda^{\alpha}\lambda^{\beta}\nabla_{\alpha}U_{\beta}^{\hphantom{\beta}\gamma I} & = & -\lambda^{\alpha}\lambda^{\beta}R_{\delta\alpha\beta}^{\hphantom{\delta\alpha\beta}\gamma}W^{\delta I}.
\end{eqnarray}
\end{subequations}All together, the constraints in \eqref{eq:hetconstraints}
imply the supergravity and super Yang-Mills equations of motion of
the heterotic background, as explained in \cite{Berkovits:2001ue}.

This work is organized as follows. Section \ref{sec:heterotic} presents
the sectorized model introduced in \cite{Jusinskas:2016qjd} for the
heterotic infinite tension string. Starting with a brief review of
Berkovits' original proposal, we will show how the BRST charge was
modified to make the sector description manifest and determine the
classical conditions for its nilpotency. In section \ref{sec:background},
we will discuss the coupling to the heterotic background. For pedagogical
reasons, we will analyze first the pure supergravity coupling and
extend the results including super Yang-Mills next, explaining in
detail how the known background constraints are obtained in the classical
analysis. Section \ref{sec:discussion} discusses the particularities
of the sectorized approach and presents some future directions to
follow. The reader is advised to go through the appendix \ref{sec:notation}
first, as the superspace conventions used here are compactly listed
there. Appendix \ref{sec:currents} contains perhaps the simplest
worldsheet model for the gauge sector with $SO(32)$ group and provides
some of the ingredients used in the main body of the text.

\section{The free heterotic string with infinite tension\label{sec:heterotic}}

\

The heterotic pure spinor string is described in the $\alpha'\to0$
limit by the chiral action
\begin{equation}
S=\frac{1}{2\pi}\int d^{2}z\{P_{a}\bar{\partial}X^{a}+p_{\alpha}\bar{\partial}\theta^{\alpha}+w_{\alpha}\bar{\partial}\lambda^{\alpha}+\bar{b}\bar{\partial}\bar{c}\}+S_{C}.\label{eq:heteroticaction}
\end{equation}
$X^{a}$ and $\theta^{\alpha}$ are the $\mathcal{N}=1$ superspace
coordinates with conjugate momenta $P_{a}$ and $p_{\alpha}$, with
$a=0,\ldots,9$ and $\alpha=1,\ldots,16$ denoting the flat vector
and spinor indices respectively. The ghost sector is represented by
the usual reparametrization ghosts, $\bar{b}$ and $\bar{c}$,
the pure spinor $\lambda^{\alpha}$, satisfying $(\lambda\gamma^{a}\lambda)=0$,
and its conjugate $w_{\alpha}$. The gamma matrices satisfy $\{\gamma^{a},\gamma^{b}\}=2\eta^{ab}$,
where $\eta^{ab}$ is the $SO(1,9)$ metric. The gauge sector is encoded
in $S_{C}$. Note that $S$ has no conformal anomaly and its energy-momentum
tensor is given by
\begin{equation}
T=-P_{a}\partial X^{a}-p_{\alpha}\partial\theta^{\alpha}-w_{\alpha}\partial\lambda^{\alpha}-\bar{b}\partial\bar{c}-\partial(\bar{b}\bar{c})+T_{C},
\end{equation}
where $T_{C}$ is the gauge sector energy-momentum tensor with central
charge $c=16$.

In \cite{Berkovits:2013xba}, the action \eqref{eq:heteroticaction}
was provided with the BRST charge
\begin{equation}
Q=\oint\{\lambda^{\alpha}[p_{\alpha}-\frac{1}{2}(\gamma^{a}\theta)_{\alpha}P_{a}]+\bar{c}T-\bar{b}\bar{c}\partial\bar{c}\}.\label{eq:BRSTnathan}
\end{equation}
However, it does not correctly describe the expected massless heterotic
spectrum, in particular it fails to reproduce the gauge transformations
of the supergravity states, which are directly related to the invariance
of the theory under general coordinate transformations.

Following the ideas of \cite{Chandia-Vallilo}, an alternative BRST
charge was proposed in \cite{Jusinskas:2016qjd} by one of the authors.
We will review this construction now.

\subsection{Review: sectorization and BRST cohomology}

\

Perhaps the first observation hinting at the inadequacy of the BRST
charge \eqref{eq:BRSTnathan} is the existence of an extra nilpotent
symmetry of the action \eqref{eq:heteroticaction}, also linear in
$\lambda^{\alpha}$, generated by
\begin{equation}
\mathcal{K}=\oint\,(\lambda\gamma_{a}\theta)[\partial X^{a}+\frac{1}{2}(\theta\gamma^{a}\partial\theta)].
\end{equation}
To consistently absorb $\mathcal{K}$ in the BRST charge, the supersymmetry
charges have to be redefined to
\begin{equation}
q_{\alpha}\equiv\oint\{p_{\alpha}+\frac{1}{2}(P_{a}-\partial X_{a})(\gamma^{a}\theta)_{\alpha}-\frac{1}{12}(\theta\gamma_{a}\partial\theta)(\gamma^{a}\theta)_{\alpha}\},
\end{equation}
which in turn brings forth the new invariants:\begin{subequations}\label{eq:flatsusyinvariants}
\begin{eqnarray}
\Pi^{a} & = & \partial X^{a}+\frac{1}{2}(\theta\gamma^{a}\partial\theta),\\
\mathcal{P}_{a} & \equiv & P_{a}-\frac{1}{2}(\theta\gamma_{a}\partial\theta),\\
d_{\alpha} & \equiv & p_{\alpha}-\frac{1}{2}P_{a}(\gamma^{a}\theta)_{\alpha}+\frac{1}{2}\Pi^{a}(\gamma_{a}\theta)_{\alpha}.
\end{eqnarray}
\end{subequations}Note that the operators $P_{a}^{\pm}$ of \cite{Jusinskas:2016qjd}
would be written here as $P_{a}^{\pm}=\mathcal{P}_{a}\pm\Pi_{a}$.
The action and its energy-momentum tensor can be expressed in terms
of the above invariants as
\begin{eqnarray}
S & = & \frac{1}{2\pi}\int d^{2}z\{\mathcal{P}_{a}\bar{\Pi}^{a}+d_{\alpha}\bar{\partial}\theta^{\alpha}+w_{\alpha}\bar{\partial}\lambda^{\alpha}+\bar{b}\bar{\partial}\bar{c}\}+S_{C}\nonumber \\
 &  & -\frac{1}{4\pi}\int d^{2}z\{\Pi^{a}(\theta\gamma_{a}\bar{\partial}\theta)-\bar{\Pi}^{a}(\theta\gamma_{a}\partial\theta)\},\label{eq:hetaction}\\
T & = & -\mathcal{P}_{a}\Pi^{a}-d_{\alpha}\partial\theta^{\alpha}-w_{\alpha}\partial\lambda^{\alpha}-\bar{b}\partial\bar{c}-\partial(\bar{b}\bar{c})+T_{C}.\label{eq:heteroticT}
\end{eqnarray}

Although not manifestly, $S$ is invariant under supersymmetry. Consider
a transformation with constant parameter $\xi^{\alpha}$, then
\begin{eqnarray}
\delta S & = & \frac{1}{4\pi}\int d^{2}z\{\bar{\Pi}^{a}(\xi\gamma_{a}\partial\theta)-\Pi^{a}(\xi\gamma_{a}\bar{\partial}\theta)\}.\nonumber \\
 & = & \frac{1}{4\pi}\int d^{2}z\{(\xi\gamma_{a}\theta)[\bar{\partial}\Pi^{a}-\partial\bar{\Pi}^{a}]\}\nonumber \\
 & = & \frac{1}{2\pi}\int d^{2}z\{(\xi\gamma_{a}\theta)(\bar{\partial}\theta\gamma^{a}\partial\theta)\},
\end{eqnarray}
Using the property $(\gamma_{\alpha\beta}^{a}\gamma_{\gamma\lambda}^{b}+\gamma_{\alpha\gamma}^{a}\gamma_{\beta\lambda}^{b}+\gamma_{\alpha\lambda}^{a}\gamma_{\gamma\beta}^{b})\eta_{ab}=0$,
the integrand in the last line can be rewritten as
\begin{equation}
(\xi\gamma_{a}\theta)(\bar{\partial}\theta\gamma^{a}\partial\theta)=\frac{1}{3}\bar{\partial}[(\xi\gamma_{a}\theta)(\theta\gamma^{a}\partial\theta)]-\frac{1}{3}\partial[(\xi\gamma_{a}\theta)(\theta\gamma^{a}\bar{\partial}\theta)],
\end{equation}
which proves the invariance of the action $S$ up to boundary terms.

We will also define the operator
\begin{equation}
\mathcal{H}\equiv-\frac{1}{2}\mathcal{P}_{a}\mathcal{P}^{a}-\frac{1}{2}\Pi_{a}\Pi^{a}+d_{\alpha}\partial\theta^{\alpha}+w_{\alpha}\partial\lambda^{\alpha}-\bar{b}\partial\bar{c}-\partial(\bar{b}\bar{c})+T_{C},\label{eq:heteroticH}
\end{equation}
which is the heterotic analogous of the generalized particle-like
Hamiltonian for the type II case of \cite{Chandia-Vallilo}. Using
these operators, it was shown in \cite{Jusinskas:2016qjd} that the
chiral action $S$ can be interpreted in terms of two sectors $(+)$
and $(-)$ with characteristic energy-momentum-like tensors
\begin{equation}
T_{\pm}\equiv\frac{1}{2}(T\pm\mathcal{H}),\label{eq:T+-}
\end{equation}
such that\begin{subequations}\label{eq:T+T-het}
\begin{eqnarray}
T_{+} & = & -\frac{1}{4}\eta^{ab}(\mathcal{P}_{a}+\Pi_{a})(\mathcal{P}_{b}+\Pi_{b})-\bar{b}\partial\bar{c}-\partial(\bar{b}\bar{c})+T_{C},\\
T_{-} & = & \frac{1}{4}\eta^{ab}(\mathcal{P}_{a}-\Pi_{a})(\mathcal{P}_{b}-\Pi_{b})-d_{\alpha}\partial\theta^{\alpha}-w_{\alpha}\partial\lambda^{\alpha}.
\end{eqnarray}
\end{subequations}

The new BRST charge makes the sectorization of the theory explicit
and is given by
\begin{equation}
Q=Q_{\lambda}+Q_{+},\label{eq:heteroticBRST}
\end{equation}
with\begin{subequations}\label{eq:partshetBRST}
\begin{eqnarray}
Q_{\lambda} & \equiv & \oint\,\lambda^{\alpha}d_{\alpha},\\
Q_{+} & \equiv & \oint\{\bar{c}T_{+}-\bar{b}\bar{c}\partial\bar{c}\}.
\end{eqnarray}
\end{subequations}$Q_{\lambda}$ is very similar to the usual (left-moving)
pure spinor BRST charge while $Q_{+}$ is composed by the familiar
BRST charge coming from the reparametrization symmetry plus an analogous
contribution with the operator $\mathcal{H}$, \emph{cf}. equation
\eqref{eq:T+-}.

The massless spectrum of the heterotic string consists of non-abelian
super Yang-Mills and $\mathcal{N}=1$ supergravity, respectively described
by the vertex operators\begin{subequations}\label{eq:heteroticvertices}
\begin{eqnarray}
U_{SYM} & = & \lambda^{\alpha}\bar{c}A_{\alpha}^{I}J_{I},\\
U_{SG} & = & \lambda^{\alpha}\bar{c}A_{\alpha}^{a}(\mathcal{P}_{a}+\Pi_{a}),
\end{eqnarray}
\end{subequations}where $J_{I}$ corresponds to (holomorphic) generators
of the $SO(32)$ or $E(8)\times E(8)$ current algebra, with $I$
denoting the adjoint representation of the gauge group. BRST-closedness
of $U_{SYM}$ and $U_{SG}$ with respect to \eqref{eq:heteroticBRST}
provides the known superfield equations of motion at the linearized
level,\begin{subequations}
\begin{eqnarray}
\gamma_{abcde}^{\alpha\beta}D_{\alpha}A_{\beta}^{I} & = & 0,\\
\gamma_{abcde}^{\alpha\beta}D_{\beta}A_{\alpha}^{f} & = & 0,\\
\partial^{b}\partial_{b}A_{\alpha}^{a}-\partial^{a}\partial_{b}A_{\alpha}^{b} & = & 0.
\end{eqnarray}
\end{subequations}The gauge transformations of the superfields, given
by\begin{subequations}
\begin{eqnarray}
\delta_{\Sigma}A_{\alpha}^{I} & = & D_{\alpha}\Sigma^{I},\\
\delta_{\Sigma}A_{\alpha}^{a} & = & D_{\alpha}\Sigma^{a}+\partial^{a}\Sigma_{\alpha},
\end{eqnarray}
\end{subequations}can be written in terms of BRST-exact expressions,
as expected. More details can be found in \cite{Jusinskas:2016qjd}.

Next, we will discuss the classical equations associated to the nilpotency
of the BRST-charge \eqref{eq:heteroticBRST} to establish the basis
for the curved background analysis of section \ref{sec:background}.

\subsection{Classical analysis\label{sub:Classical-analysis}}

\

In order to determine the classical conditions to be imposed on the
background, it might be useful to understand their meaning in the
flat case. Recall that the heterotic action can be cast as
\begin{eqnarray}
S & = & \frac{1}{2\pi}\int d^{2}z\{\mathcal{P}_{a}\bar{\Pi}^{a}+d_{\alpha}\bar{\partial}\theta^{\alpha}+w_{\alpha}\bar{\partial}\lambda^{\alpha}+\bar{b}\bar{\partial}\bar{c}\}+S_{C}\nonumber \\
 &  & -\frac{1}{4\pi}\int d^{2}z\{\Pi^{a}(\theta\gamma_{a}\bar{\partial}\theta)-\bar{\Pi}^{a}(\theta\gamma_{a}\partial\theta)\},
\end{eqnarray}
with $\mathcal{P}_{a}$ and $d_{\alpha}$ being supersymmetric invariants
defined in terms of the conjugate momenta of $X^{a}$ and $\theta^{\alpha}$
respectively, \emph{cf}. equation \eqref{eq:flatsusyinvariants}.
It is convenient, however, to treat them as independent variables.
The above action is just one step behind the curved space one that
we will define in the next section.

The BRST symmetry is described by the charge displayed in \eqref{eq:heteroticBRST}.
To compute the classical BRST transformations of the worldsheet variables,
we will rewrite $Q$ in terms of the fields $\{X^{a},\theta^{\alpha},\lambda^{\alpha},\bar{c}\}$,
collectively denoted by $\phi$, and their canonical conjugates, which
are given in terms of $\{\mathcal{P}_{a},d_{\alpha},w_{\alpha},\bar{b}\}$.
The latter will be denoted by $\hat{P}_{\phi}$ and are usually defined
with respect to $\tau$, the worldsheet time. We will use the Minkowski
parametrization with $z=\sigma-\tau$ and $\bar{z}=\sigma+\tau$,
where $\sigma\in[0,2\pi)$ denotes the spatial coordinate. The derivatives
can then be cast as
\begin{equation}
\begin{array}{cc}
\partial=\frac{1}{2}(\partial_{\sigma}-\partial_{\tau}), & \bar{\partial}=\frac{1}{2}(\partial_{\sigma}+\partial_{\tau}).\end{array}
\end{equation}
With this convention, the canonical momenta will be defined to be
\begin{equation}
\hat{P}[\phi]\equiv2\pi\Bigg(\frac{\delta S}{\delta(\bar{\partial}\phi)}-\frac{\delta S}{\delta(\partial\phi)}\Bigg),\label{eq:defcanonical}
\end{equation}
leading to the following identifications:\begin{subequations}
\begin{eqnarray}
\hat{P}[X^{a}] & = & \mathcal{P}_{a}+\frac{1}{2}(\theta\gamma_{a}\partial_{\sigma}\theta),\nonumber \\
 & \equiv & \hat{P}_{a}\label{eq:conjugateX}\\
\hat{P}[\theta^{\alpha}] & = & -d_{\alpha}-\frac{1}{2}(\mathcal{P}_{a}-\partial_{\sigma}X_{a})(\gamma^{a}\theta)_{\alpha},\nonumber \\
 & \equiv & \hat{P}_{\alpha}\label{eq:conjugatetheta}\\
\hat{P}[\lambda^{\alpha}] & = & w_{\alpha},\\
\hat{P}[\bar{c}] & = & -\bar{b}.
\end{eqnarray}
\end{subequations}

The fundamental Poisson brackets are simply given by
\begin{equation}
\left\{ \hat{P}[\phi'(\sigma')],\phi(\sigma)\right\} _{P.B.}=-\delta_{\phi,\phi'}\delta(\sigma-\sigma').
\end{equation}
Therefore, the BRST transformations of the worldsheet fields are easily
computed when the BRST charge is written in terms of $\hat{P}$ and
$\phi$. For example, $Q_{\lambda}$ in \eqref{eq:partshetBRST} is
expressed as
\begin{equation}
Q_{\lambda}=\oint d\sigma\lambda^{\alpha}[-\hat{P}_{\alpha}-\frac{1}{2}\hat{P}_{a}(\gamma^{a}\theta)_{\alpha}+\frac{1}{2}\Pi_{\sigma}^{a}(\gamma_{a}\theta)_{\alpha}].
\end{equation}

Concerning the nilpotency of the BRST charge $Q$, it can be stated
as
\begin{equation}
Q_{\lambda}^{2}+\{Q_{\lambda},Q_{+}\}+Q_{+}^{2}=0.
\end{equation}
Because $Q_{\lambda}$ is independent of the reparametrization ghosts,
each term in the equation above should vanish separately.
Therefore, following the classical construction just presented, it is easy
to demonstrate that $Q$ is nilpotent if and only if\begin{subequations}\label{eq:nilpotency-conditions}
\begin{eqnarray}
Q_{\lambda}^{2} & = & 0,\label{eq:nilpotencylambda2}\\
\{\lambda^{\alpha}d_{\alpha}(\sigma'),T_{+}(\sigma)\}_{P.B.} & = & 0,\label{eq:nilpotencyTlambda}\\
\{T_{+}(\sigma'),T_{+}(\sigma)\}_{P.B.} & = & 2T_{+}\,\partial_{\sigma}\delta(\sigma'-\sigma)+\partial_{\sigma}T_{+}\,\delta(\sigma'-\sigma).\label{eq:nilpotencyTT}
\end{eqnarray}
\end{subequations}

In flat space, it is straightforward to see that all these relations
are satisfied. In the next section they will be our guidelines for
nontrivial backgrounds. The difference then will be how the background
manifests itself in the definition of the conjugate momenta, in particular
\eqref{eq:conjugateX} and \eqref{eq:conjugatetheta}, which contain
the fundamental ingredients of the BRST charge, $\mathcal{P}_{a}$
and $d_{\alpha}$.

\section{Classical consistency of the heterotic background\label{sec:background}}

\

In this section we will show how the nilpotency conditions discussed
above ultimately impose constraints on the heterotic background, providing
the expected supergravity and super Yang-Mills equations of motion
in superspace detailedly presented in \cite{Berkovits:2001ue} for
the pure spinor superstring.

After understanding how the infinite tension string couples to the
heterotic background, we will be able to build the operator set necessary
for our analysis. The supergravity sector is presented alone beforehand
for two reasons. First, to the best of our knowledge, there is no
good description for $\mathcal{N}=1$ (heterotic) supergravity in
any ambitwistor string so far. So this will be a good test for the
modifications discussed in \cite{Jusinskas:2016qjd} for the sectorized
string. Second, the generalization from flat space is straightforward
and it will help establish the curved superspace language that is
extensively used. Next, we will turn on the super Yang-Mills background
and extend the results.

\subsection{Supergravity background and constraints}

\

The curved superspace generalization of \eqref{eq:hetaction} is given
by
\begin{equation}
S=\frac{1}{2\pi}\int d^{2}z\{\mathcal{P}_{a}\bar{\Pi}^{a}+d_{\alpha}\bar{\Pi}^{\alpha}-\Pi^{A}\bar{\Pi}^{B}B_{BA}+w_{\alpha}\bar{\nabla}\lambda^{\alpha}+\bar{b}\bar{\partial}\bar{c}\}+S_{C}.\label{eq:hetaction_curved}
\end{equation}
The vielbein $E_{M}^{\hphantom{M}A}$, and the Lorentz connection
$\Omega_{AB}^{\hphantom{AB}C}$, enter the action through the generalized
superspace invariants and the covariant derivative\footnote{Due to the pure spinor constraint the action has a gauge symmetry
with parameter $\varphi_{a}$ of the form $\delta_{\varphi}w_{\alpha}=\varphi_{a}(\gamma^{a}\lambda)_{\alpha}$.
Therefore, to work only with gauge invariant quantities we must impose
$\lambda^{\alpha}(\gamma^{a}\lambda)_{\beta}\Omega_{A \alpha}^{\hphantom{A \alpha}\beta}=0$.}, given by\begin{subequations}
\begin{eqnarray}
\bar{\Pi}^{A} & = & \bar{\partial}Z^{M}E_{M}^{\hphantom{M}A},\\
\bar{\nabla}\lambda^{\alpha} & = & \bar{\partial}\lambda^{\alpha}+\lambda^{\beta}\bar{\Pi}^{A}\Omega_{A\beta}^{\hphantom{A\beta}\alpha},
\end{eqnarray}
\end{subequations}and analogous expressions for $\Pi^{A}$ and $\nabla\lambda^{\alpha}$,
where $Z^{M}$ denotes the curved $\mathcal{N}=1$ superspace coordinates
$X^{m}$ and $\theta^{\mu}$. The curved vector and spinor indices
are being respectively denoted by $m=0,\ldots,9$ and $\mu=1,\ldots,16$.
Notice that in this language the supermetric $G_{MN}$ is written
in terms of the flat metric as $G_{MN}=E_{M}^{\hphantom{M}a}E_{N}^{\hphantom{N}b}\eta_{ab}$.
The coupling with the Kalb-Ramond superfield can be easily written
with explicit curved space indices,
\begin{eqnarray}
S_{B} & = & -\frac{1}{2\pi}\int d^{2}z\{\Pi^{A}\bar{\Pi}^{B}B_{BA}\}\nonumber \\
 & = & -\frac{1}{2\pi}\int d^{2}z\{\partial Z^{M}\bar{\partial}Z^{N}B_{NM}\},
\end{eqnarray}
with
\begin{equation}
B_{AB}=(-1)^{A(B+N)}E_{B}^{\hphantom{B}N}E_{A}^{\hphantom{A}M}B_{MN}.
\end{equation}
This form is more suitable to show the gauge invariance of the action
with respect to the transformations $\delta B_{MN}=\partial_{M}\Sigma_{N}-(-1)^{MN}\partial_{N}\Sigma_{M}$.
More details on the conventions used here can be found in Appendix
\ref{sec:notation}. Concerning the dilaton superfield, it plays no
role in the classical description and this can be seen from the fact
that its coupling to the action naively vanishes in the $\alpha'\to0$ limit. 

Following the analysis of subsection \ref{sub:Classical-analysis},
$\mathcal{P}_{a}$ and $d_{\alpha}$ can be viewed as independent
objects invariant under supesymmetry, and the flat space limit of
$S$ is recovered when we express them in terms of regular variables,
\emph{cf.} \eqref{eq:flatsusyinvariants}, together with the non-vanishing
components of $E$ and $B$ in that limit:
\begin{equation}
\begin{array}{cc}
E_{m}^{\hphantom{m}a}=\delta_{m}^{a}, & E_{\mu}^{\hphantom{\mu}a}=-\frac{1}{2}(\gamma^{a}\theta)_{\mu}\\
E_{\mu}^{\hphantom{\mu}\alpha}=\delta_{\mu}^{\alpha}, & B_{m\mu}=-B_{\mu m}=\frac{1}{2}E_{m}^{\hphantom{m}a}(\gamma_{a}\theta)_{\mu}.
\end{array}
\end{equation}

The energy-momentum tensor of the curved space action is given by
\begin{equation}
T=-\mathcal{P}_{a}\Pi^{a}-d_{\alpha}\Pi^{\alpha}-w_{\alpha}\nabla\lambda^{\alpha}-\bar{b}\partial\bar{c}-\partial(\bar{b}\bar{c})+T_{C},\label{eq:SUGRAT}
\end{equation}
and the curved version of $\mathcal{H}$ in \eqref{eq:heteroticH}
is simply
\begin{equation}
\mathcal{H}=-\frac{1}{2}\mathcal{P}_{a}\mathcal{P}^{a}-\frac{1}{2}\Pi^{a}\Pi_{a}+d_{\alpha}\Pi^{\alpha}+w_{\alpha}\nabla\lambda^{\alpha}-\bar{b}\partial\bar{c}-\partial(\bar{b}\bar{c})+T_{C}.\label{eq:SUGRAH}
\end{equation}

The BRST-charge in the curved background has the same structure of
\eqref{eq:heteroticBRST} and the presence of the background can be
seen through the canonical conjugates of the superspace coordinates
$Z^{M}$, denoted by $\hat{P}_{M}$. Using the definition \eqref{eq:defcanonical},
one obtains
\begin{equation}
\hat{P}_{M}=E_{M}^{\hphantom{M}a}\mathcal{P}_{a}-E_{M}^{\hphantom{M}\alpha}d_{\alpha}+\partial_{\sigma}Z^{N}B_{NM}+w_{\alpha}\lambda^{\beta}\Omega_{M\beta}^{\hphantom{M\beta}\alpha},\label{eq:curvedcanonical}
\end{equation}
which enables us to rewrite $\mathcal{P}_{a}$ and $d_{\alpha}$ as\begin{subequations}\label{eq:curvedPdconjugates}
\begin{eqnarray}
d_{\alpha} & = & -E_{\alpha}^{\hphantom{\alpha}M}\hat{P}_{M}+(\Pi^{A}+\bar{\Pi}^{A})B_{A\alpha}+w_{\gamma}\lambda^{\beta}\Omega_{\alpha\beta}^{\hphantom{\alpha\beta}\gamma},\label{eq:dcurved}\\
\mathcal{P}_{a} & = & E_{a}^{\hphantom{\alpha}M}\hat{P}_{M}-(\Pi^{A}+\bar{\Pi}^{A})B_{Aa}-w_{\gamma}\lambda^{\beta}\Omega_{a\beta}^{\hphantom{a\beta}\gamma}.\label{eq:Pcurved}
\end{eqnarray}
\end{subequations}To go from \eqref{eq:curvedcanonical} to \eqref{eq:curvedPdconjugates},
we have used the inverse vielbein $E_{A}^{\hphantom{A}M}$, such that
$E_{A}^{\hphantom{A}M}E_{M}^{\hphantom{M}B}=\delta_{A}^{B}$ and $E_{M}^{\hphantom{N}A}E_{A}^{\hphantom{A}N}=\delta_{M}^{N}$. 

For the BRST-charge to be nilpotent, and thus well-defined as such,
the background superfields need to satisfy a number of constraints.
To find these constraints, we begin by computing the transformations
of the worldsheet fields under the action of $Q_{\lambda}$. Using
the graded Poisson brackets
\begin{eqnarray}
\{\hat{P}_{M}(\sigma'),Z^{N}(\sigma)\}_{P.B.} & = & -\delta_{M}^{N}\delta(\sigma-\sigma^{\prime}),\label{eq:canonicalZP}\\
\{w_{\alpha}(\sigma'),\lambda^{\beta}(\sigma)\}_{P.B.} & = & -\delta_{\alpha}^{\beta}\delta(\sigma-\sigma^{\prime}),\label{eq:poissonlambda}
\end{eqnarray}
we obtain the following transformations\footnote{For a reasonably detailed exposition of similar calculations, see
\cite{Chandia:2006ix}.} \begin{subequations}\label{eq:curvedbrsttransf}
\begin{eqnarray}
\delta\lambda^{\alpha} & = & -\lambda^{\beta}\Lambda_{\beta}^{\hphantom{\beta}\alpha},\\
\delta w_{\alpha} & = & \Lambda_{\alpha}^{\hphantom{\alpha}\beta}w_{\beta}+\epsilon d_{\alpha},\\
\delta\Pi^{a} & = & -\Pi^{b}\Lambda_{b}^{\hphantom{b}a}-\epsilon\lambda^{\alpha}\Pi^{A}T_{A\alpha}^{\hphantom{A\alpha}a},\\
\delta\Pi^{\alpha} & = & -\Pi^{\beta}\Lambda_{\beta}^{\hphantom{\beta}\alpha}+\epsilon\nabla\lambda^{\alpha}-\epsilon\lambda^{\beta}\Pi^{A}T_{A\beta}^{\hphantom{A\beta}\alpha},\\
\delta\mathcal{P}_{a} & = & \Lambda_{a}^{\hphantom{a}b}\mathcal{P}_{b}-\epsilon\lambda^{\beta}T_{\beta a}^{\hphantom{\beta a}b}\mathcal{P}_{b}+\epsilon\lambda^{\gamma}T_{\gamma a}^{\hphantom{\gamma a}\beta}d_{\beta}+\epsilon\lambda^{\beta}\Pi^{A}H_{A\beta a}-\epsilon\lambda^{\gamma}\lambda^{\beta}w_{\delta}R_{\beta a\gamma}^{\hphantom{\beta a\gamma}\delta},\\
\delta d_{\alpha} & = & \Lambda_{\alpha}^{\hphantom{\alpha}\beta}d_{\beta}+\epsilon\lambda^{\beta}T_{\beta\alpha}^{\hphantom{\beta\alpha}b}\mathcal{P}_{b}-\epsilon\lambda^{\gamma}T_{\gamma\alpha}^{\hphantom{\gamma\alpha}\beta}d_{\beta}-\epsilon\lambda^{\beta}\Pi^{A}H_{A\beta\alpha}+\epsilon\lambda^{\gamma}\lambda^{\beta}w_{\delta}R_{\beta\alpha\gamma}^{\hphantom{\beta\alpha\gamma}\delta},\\
\delta\Omega_{\alpha}^{\hphantom{\alpha}\beta} & = & \nabla\Lambda_{\alpha}^{\hphantom{\alpha}\beta}-\epsilon\lambda^{\gamma}\Pi^{A}R_{A\gamma\alpha}^{\hphantom{A\gamma\alpha}\beta}.
\end{eqnarray}
\end{subequations}Here $\epsilon$ is a constant anticommuting parameter
and we have defined
\begin{equation}
\begin{array}{cc}
\Lambda_{A}^{\hphantom{A}B}\equiv\epsilon\lambda^{\alpha}\Omega_{\alpha A}^{\hphantom{\alpha A}B}, & \Omega_{\alpha}^{\hphantom{\alpha}\beta}\equiv\Pi^{A}\Omega_{A\alpha}^{\hphantom{A\alpha}\beta}.\end{array}
\end{equation}

Now we can compute the transformation of $\lambda^{\alpha}d_{\alpha}$,
whose vanishing is equivalent to the first condition displayed in
\eqref{eq:nilpotency-conditions}:
\begin{equation}
\delta(\lambda^{\alpha}d_{\alpha})=\epsilon\lambda^{\alpha}\lambda^{\beta}[T_{\alpha\beta}^{\hphantom{\alpha\beta}a}\mathcal{P}_{a}-T_{\alpha\beta}^{\hphantom{\alpha\beta}\gamma}d_{\gamma}-\Pi^{A}H_{A\alpha\beta}+w_{\delta}\lambda^{\gamma}R_{\alpha\beta\gamma}^{\hphantom{\alpha\beta\gamma}\delta}].\label{eq:SUGRAdeltaQlambda}
\end{equation}
 Hence the first set of constraints required for the nilpotency of
$Q$ is:
\begin{equation}
\lambda^{\alpha}\lambda^{\beta}T_{\alpha\beta}^{\hphantom{\alpha\beta}A}=\lambda^{\alpha}\lambda^{\beta}H_{A\alpha\beta}=\lambda^{\alpha}\lambda^{\beta}\lambda^{\gamma}R_{\alpha\beta\gamma}^{\hphantom{\alpha\beta\gamma}\delta}=0.\label{eq:nilpotencySUGRAconstraints}
\end{equation}

Nilpotency of the BRST charge also requires $\delta T_{+}$ to vanish.
This is just another way of stating the condition \eqref{eq:nilpotencyTlambda}.
The operator $T_{+}$ is obtained from the definition \eqref{eq:T+-}
and the curved versions of $T$ and $\mathcal{H}$, respectively \eqref{eq:SUGRAT}
and \eqref{eq:SUGRAH}. It can be cast as
\begin{equation}
T_{+}=-\frac{1}{4}\eta^{ab}(\mathcal{P}_{a}+\Pi_{a})(\mathcal{P}_{b}+\Pi_{b})-\bar{b}\partial\bar{c}-\partial(\bar{b}\bar{c})+T_{C}.\label{eq:T+curved}
\end{equation}
Now, to compute $\delta T_{+}$ we just have to use the transformations
of $\mathcal{P}_{a}$ and $\Pi^{a}$ in \eqref{eq:curvedbrsttransf}
and the result is
\begin{eqnarray}
\delta T_{+} & = & -\frac{1}{2}\epsilon(\mathcal{P}^{a}+\Pi^{a})\lambda^{\alpha}d_{\beta}T_{\alpha a}^{\hphantom{\alpha a}\beta}+\frac{1}{2}\epsilon(\mathcal{P}^{a}\mathcal{P}^{b}-\Pi^{a}\Pi^{b})\lambda^{\alpha}T_{\alpha ab}-\frac{1}{2}\epsilon\lambda^{\alpha}\mathcal{P}^{a}\Pi^{b}H_{\alpha ab}\nonumber \\
 &  & +\frac{1}{2}\epsilon(\mathcal{P}^{a}+\Pi^{a})\lambda^{\alpha}\Pi^{\beta}(T_{\alpha\beta a}-H_{\alpha\beta a})+\frac{1}{2}\epsilon(\mathcal{P}^{a}+\Pi^{a})\lambda^{\alpha}\lambda^{\beta}w_{\gamma}R_{\beta a\alpha}^{\hphantom{\beta a\alpha}\gamma}.
\end{eqnarray}
For this expression to vanish, we need to impose another set constraints:
\begin{equation}
T_{\alpha a}^{\hphantom{\alpha a}\beta}=T_{\alpha(ab)}=T_{\alpha\beta b}-H_{\alpha\beta b}=H_{ab\alpha}=\lambda^{\alpha}\lambda^{\beta}R_{\alpha a\beta}^{\hphantom{\alpha a\beta}\gamma}=0.\label{eq:holoSUGRAconstraints}
\end{equation}
In the usual pure spinor superstring, this set comes from the holomorphicity
of the BRST charge \cite{Berkovits:2001ue}.

Finally, rewriting $T_{+}$ in terms of the canonical conjugates and
using the Poisson brackets of \eqref{eq:canonicalZP} together with
\begin{eqnarray}
\{\bar{b}(\sigma'),\bar{c}(\sigma)\}_{P.B.} & = & \delta(\sigma'-\sigma),\label{eq:poissonbc}\\
\{T_{C}(\sigma'),T_{C}(\sigma)\}_{P.B.} & = & 2T_{C}\,\partial_{\sigma}\delta(\sigma'-\sigma)+\partial_{\sigma}T_{C}\,\delta(\sigma'-\sigma),\label{eq:poissonTc}
\end{eqnarray}
we can show that
\begin{equation}
\{T_{+}(\sigma'),T_{+}(\sigma)\}_{P.B.}=2T_{+}\,\partial_{\sigma}\delta(\sigma'-\sigma)+\partial_{\sigma}T_{+}\,\delta(\sigma'-\sigma).
\end{equation}
Therefore, the three conditions of \eqref{eq:nilpotency-conditions}
for classical nilpotency of the BRST charge in a supergravity background
are all satisfied provided the constraints displayed in \eqref{eq:nilpotencySUGRAconstraints}
and \eqref{eq:holoSUGRAconstraints}.

\subsection{Turning on the super Yang-Mills background}

\ 

In order to find the remaining heterotic background constraints, we
need to consider the case in which the super Yang-Mills fields are
present. We will introduce the minimal coupling between the gauge
potential $A_{M}^{I}$ and the currents $J_{I}$, such that the action
has the form
\begin{eqnarray}
S & = & \frac{1}{2\pi}\int d^{2}z\{\mathcal{P}_{a}\bar{\Pi}^{a}+d_{\alpha}\bar{\Pi}^{\alpha}-\Pi^{A}\bar{\Pi}^{B}B_{BA}+\bar{A}^{I}J_{I}+w_{\alpha}\bar{\nabla}\lambda^{\alpha}+\bar{b}\bar{\partial}\bar{c}\}+S_{C},\label{eq:hetaction_SYM}
\end{eqnarray}
where $\bar{A}^{I}\equiv\bar{\partial}Z^{M}A_{M}^{I}$.
Note this is equivalent to replacing $\mathcal{P}_{a}\to\mathcal{P}_{a}+A_{a}^{I}J_{I}$
and $d_{\alpha}\to d_{\alpha}-A_{\alpha}^{I}J_{I}$ in the action
\ref{eq:hetaction_curved}, with $A_{A}^{I}=E_{A}^{\hphantom{A}M}A_{M}^{I}$.

The gauge invariance of the action \eqref{eq:hetaction_SYM} with
respect to the super Yang-Mills background is straightforward to demonstrate.
Consider the gauge transformations with superparameter $\Sigma^{I}$,
\begin{equation}
\delta_{\Sigma}A_{M}^{I}=\partial_{M}\Sigma^{I}+[A_{M},\Sigma]^{I},\label{eq:SYMgauge}
\end{equation}
where
\begin{equation}
[A_{M},\Sigma]^{I}=f_{JK}^{\hphantom{JK}I}A_{M}^{J}\Sigma^{K},
\end{equation}
and $f_{JK}^{\hphantom{JK}I}$ denotes the structure constants of
the gauge group. Substituting the transformation \eqref{eq:SYMgauge}
in the variation of the action and integrating by parts, one obtains
\begin{equation}
\delta_{\Sigma}S=-\frac{1}{2\pi}\int d^{2}z(\bar{\partial}J_{I}+f_{IJ}^{\hphantom{IJ}K}\bar{A}^{J}J_{K})\Sigma^{I}.
\end{equation}
The expression inside the parentheses is just the equation of motion
for the current $J_{I}$ in the presence of the super Yang-Mills source
and thus vanishes at the classical level. More details on this construction
can be found in appendix \ref{sec:currents} where an explicit realization
of the gauge sector is given for the group $SO(32)$ in terms of worldsheet
fermions. For convenience, we can introduce the super field strength
of $A_{M}^{I}$, given by
\begin{eqnarray}
F_{MN}^{I} & = & \partial_{M}A_{N}^{I}-(-1)^{MN}\partial_{N}A_{M}^{I}+f_{JK}^{\hphantom{JK}I}A_{M}^{J}A_{N}^{K},\label{eq:superfieldstrength}
\end{eqnarray}
which transforms covariantly under \eqref{eq:SYMgauge},
\begin{equation}
\delta_{\Sigma}F_{MN}^{I}=[F_{MN},\Sigma]^{I}.
\end{equation}

The coupling to the super Yang-Mills background changes the energy-momentum
tensor to
\begin{equation}
T=-\mathcal{P}_{a}\Pi^{a}-d_{\alpha}\Pi^{\alpha}-w_{\alpha}\nabla\lambda^{\alpha}-\bar{b}\partial\bar{c}-\partial(\bar{b}\bar{c})+T_{C}-\Pi^{A}A_{A}^{I}J_{I},\label{eq:SYMT}
\end{equation}
and we also expect the operator $\mathcal{H}$ to be modified accordingly.
It was suggested in \cite{Chandia-Vallilo} that fluctuations of the
background would be manifested through $\mathcal{H}$. Therefore,
inspired by the superstring integrated vertex, we propose
\begin{eqnarray}
\mathcal{H} & = & -\frac{1}{2}\mathcal{P}_{a}\mathcal{P}^{a}-\frac{1}{2}\Pi^{a}\Pi_{a}+d_{\alpha}\Pi^{\alpha}+w_{\alpha}\nabla\lambda^{\alpha}-\bar{b}\partial\bar{c}-\partial(\bar{b}\bar{c})+T_{C}\nonumber \\
 &  & -\Pi^{A}A_{A}^{I}J_{I}-d_{\alpha}W^{\alpha I}J_{I}-\lambda^{\alpha}w_{\beta}U_{\alpha}^{\hphantom{\alpha}\beta I}J_{I},\label{eq:SYMH}
\end{eqnarray}
where $W^{\alpha I}$ and $U_{\alpha}^{\hphantom{\alpha}\beta I}$
are background superfields\footnote{As before, we must impose $\lambda^{\alpha}(\gamma^{a}\lambda)_{\beta}U_{\alpha}^{\hphantom{\alpha}\beta I}=0$ in order to respect the gauge invariance implied by the pure spinor constraint.} which will be related to \eqref{eq:superfieldstrength}. Again, using
\eqref{eq:T+-}, $T_{+}$ can be cast as
\begin{eqnarray}
T_{+} & = & -\frac{1}{4}\eta^{ab}(\mathcal{P}_{a}+\Pi_{a})(\mathcal{P}_{b}+\Pi_{b})-\bar{b}\partial\bar{c}-\partial(\bar{b}\bar{c})+T_{C}\nonumber \\
 &  & -\Pi^{A}A_{A}^{I}J_{I}-\frac{1}{2}d_{\alpha}W^{\alpha I}J_{I}-\frac{1}{2}\lambda^{\alpha}w_{\beta}U_{\alpha}^{\hphantom{\alpha}\beta I}J_{I},\label{eq:SYMT+}
\end{eqnarray}
and we now have all the ingredients to analyze the BRST symmetry in
this background.

The modification of the action entails a change in the classical BRST
transformations of the worldsheet fields. This is clearly seen from
the canonical conjugates of the superspace coordinates, which now
have a linear dependence on the gauge field:
\begin{equation}
\hat{P}_{M}=E_{M}^{\hphantom{M}a}\mathcal{P}_{a}-E_{M}^{\hphantom{M}\alpha}d_{\alpha}+\partial_{\sigma}Z^{N}B_{NM}+w_{\alpha}\lambda^{\beta}\Omega_{M\beta}^{\hphantom{M\beta}\alpha}+A_{M}^{I}J_{I}.\label{eq:curvedSYMcanonical}
\end{equation}
To compute these transformations we will use the fundamental brackets
of \eqref{eq:canonicalZP}, \eqref{eq:poissonlambda}, \eqref{eq:poissonbc}
and \eqref{eq:poissonTc}, together with\begin{subequations}
\begin{eqnarray}
\{T_{C}(\sigma'),J_{I}(\sigma)\}_{P.B.} & = & J_{I}\partial_{\sigma}\delta(\sigma-\sigma')+\partial_{\sigma}J_{I}\delta(\sigma'-\sigma),\\
\{J_{I}(\sigma'),J_{J}(\sigma)\}_{P.B.} & = & f_{IJ}^{\hphantom{IJ}K}J_{K}\delta(\sigma'-\sigma),
\end{eqnarray}
\end{subequations}which are derived in Appendix \ref{sec:currents}.

Considering first $Q_{\lambda}$, it is clear that most of the transformations
displayed in \eqref{eq:curvedbrsttransf} remain unchanged, except
for $\delta\mathcal{P}_{a}$ and $\delta d_{\alpha}$, which are now
given by\begin{subequations}
\begin{eqnarray}
\delta\mathcal{P}_{a} & = & \Lambda_{a}^{\hphantom{a}b}\mathcal{P}_{b}-\epsilon\lambda^{\beta}T_{\beta a}^{\hphantom{\beta a}b}\mathcal{P}_{b}+\epsilon\lambda^{\gamma}T_{\gamma a}^{\hphantom{\gamma a}\beta}d_{\beta}\nonumber \\
 &  & +\epsilon\lambda^{\alpha}\Pi^{A}H_{A\alpha a}-\epsilon\lambda^{\gamma}\lambda^{\beta}w_{\delta}R_{\beta a\gamma}^{\hphantom{\beta a\gamma}\delta}-\epsilon\lambda^{\beta}F_{\beta a}^{I}J_{I},\\
\delta d_{\alpha} & = & \Lambda_{\alpha}^{\hphantom{\alpha}\beta}d_{\beta}+\epsilon\lambda^{\beta}T_{\beta\alpha}^{\hphantom{\beta\alpha}b}\mathcal{P}_{b}-\epsilon\lambda^{\gamma}T_{\gamma\alpha}^{\hphantom{\gamma\alpha}\beta}d_{\beta}\nonumber \\
 &  & -\epsilon\lambda^{\beta}\Pi^{A}H_{A\beta\alpha}+\epsilon\lambda^{\gamma}\lambda^{\beta}w_{\delta}R_{\beta\alpha\gamma}^{\hphantom{\beta\alpha\gamma}\delta}+\epsilon\lambda^{\beta}F_{\beta\alpha}^{I}J_{I},
\end{eqnarray}
\end{subequations}where
\begin{equation}
F_{AB}^{I}=(-1)^{A(B+N)}E_{B}^{\hphantom{B}N}E_{A}^{\hphantom{A}M}F_{MN}^{I}.
\end{equation}
We can now easily compute the transformation of $\lambda^{\alpha}d_{\alpha}$
and the result is
\begin{equation}
\delta(\lambda^{\alpha}d_{\alpha})=\epsilon\lambda^{\alpha}\lambda^{\beta}[T_{\alpha\beta}^{\hphantom{\alpha\beta}a}\mathcal{P}_{a}-T_{\alpha\beta}^{\hphantom{\alpha\beta}\gamma}d_{\gamma}-\Pi^{A}H_{A\alpha\beta}+w_{\delta}\lambda^{\gamma}R_{\alpha\beta\gamma}^{\hphantom{\alpha\beta\gamma}\delta}+F_{\alpha\beta}^{I}J_{I}].
\end{equation}
Thus, together with the constraints displayed in \eqref{eq:nilpotencySUGRAconstraints}
we also need to impose 
\begin{equation}
\lambda^{\alpha}\lambda^{\beta}F_{\alpha\beta}^{I}=0,\label{eq:nilpotencySYM}
\end{equation}
in order to satisfy the first nilpotency condition, \emph{cf}. equation
\eqref{eq:nilpotencylambda2}.

Next, to compute $\delta T_{+}$ and evaluate the condition \eqref{eq:nilpotencyTlambda}
it is worth noting that $T_{C}$ and $J_{I}$ now have nonvanishing
transformations with respect to $Q_{\lambda}$, given by\begin{subequations}
\begin{eqnarray}
\delta T_{C} & = & J_{I}\partial\Sigma^{I},\\
\delta J_{I} & = & -f_{IJ}^{\hphantom{IJ}K}\Sigma^{J}J_{K},
\end{eqnarray}
\end{subequations}where we have defined the gauge-like parameter
$\Sigma^{I}\equiv\epsilon\lambda^{\alpha}A_{\alpha}^{I}$. The introduction
of $\Sigma^{I}$ is convenient when we look at the variations of the
background superfields,\begin{subequations}
\begin{eqnarray}
\delta A^{I} & = & \nabla\Sigma^{I}-\epsilon\lambda^{\alpha}\Pi^{A}F_{A\alpha}^{I},\\
\delta W^{\alpha I} & = & -W^{\beta I}\Lambda_{\beta}^{\hphantom{\beta}\alpha}-[\Sigma,W^{\alpha}]^{I}+\epsilon\lambda^{\beta}\nabla_{\beta}W^{\alpha I},\\
\delta U_{\alpha}^{\hphantom{\alpha}\beta I} & = & \Lambda_{\alpha}^{\hphantom{\alpha}\gamma}U_{\gamma}^{\hphantom{\gamma}\beta I}-U_{\alpha}^{\hphantom{\alpha}\gamma I}\Lambda_{\gamma}^{\hphantom{\gamma}\beta}-[\Sigma,U_{\alpha}^{\hphantom{\alpha}\beta}]^{I}+\epsilon\lambda^{\gamma}\nabla_{\gamma}U_{\alpha}^{\hphantom{\alpha}\beta I},
\end{eqnarray}
\end{subequations}which can be interpreted in terms of a gauge-like
transformation with parameter $\Sigma$, a Lorentz-like transformation
with parameter $\Lambda$, and a superspace translation, with $\nabla_{\alpha}$
denoting the covariant derivative with respect to the local symmetries,
\emph{e.g.}
\begin{equation}
\nabla_{\beta}W^{\alpha I}=D_{\beta}W^{\alpha I}+\Omega_{\beta\gamma}^{\hphantom{\beta\gamma}\alpha}W^{\gamma I}+[A_{\beta},W^{\alpha}]^{I}.
\end{equation}

Gathering all these results and using the supergravity constraints
of \eqref{eq:holoSUGRAconstraints}, we obtain
\begin{eqnarray}
\delta T_{+} & = & \frac{1}{2}\epsilon\lambda^{\alpha}(\mathcal{P}^{a}-\Pi^{a})[F_{\alpha a}^{I}-T_{\alpha\beta a}W^{\beta I}]J_{I}-\frac{1}{2}\epsilon\lambda^{\alpha}\lambda^{\beta}w_{\gamma}[\nabla_{\alpha}U_{\beta}^{\hphantom{\beta}\gamma I}+R_{\delta\alpha\beta}^{\hphantom{\delta\alpha\beta}\gamma}W^{\delta I}]J_{I}\nonumber \\
 &  & +\epsilon\lambda^{\alpha}\Pi^{\beta}[F_{\alpha\beta}^{I}+\frac{1}{2}H_{\alpha\beta\gamma}W^{\gamma I}]J_{I}+\frac{1}{2}\epsilon\lambda^{\alpha}d_{\beta}[\nabla_{\alpha}W^{\beta I}-T_{\alpha\gamma}^{\hphantom{\alpha\gamma}\beta}W^{\gamma I}-U_{\alpha}^{\hphantom{\alpha}\beta I}]J_{I}\nonumber \\
 &  & -\frac{1}{2}\epsilon\lambda^{\alpha}F_{\alpha\beta}^{I}W^{\beta J}J_{I}J_{J},\label{newvariationH}
\end{eqnarray}
Hence, we have to further impose the following constraints:\begin{subequations}\label{eq:holoSYM}
\begin{eqnarray}
F_{\alpha a}^{I} & = & T_{\alpha\beta a}W^{\beta I},\\
\nabla_{\alpha}W^{\beta I}-T_{\alpha\gamma}^{\hphantom{\alpha\gamma}\beta}W^{\gamma I} & = & U_{\alpha}^{\hphantom{\alpha}\beta I},\\
F_{\alpha\beta}^{I} & = & \frac{1}{2}W^{\gamma I}H_{\alpha\beta\gamma},\label{eq:F=00003DWH}\\
\lambda^{\alpha}\lambda^{\beta}\nabla_{\alpha}U_{\beta}^{\hphantom{\beta}\gamma I} & = & -\lambda^{\alpha}\lambda^{\beta}R_{\delta\alpha\beta}^{\hphantom{\delta\alpha\beta}\gamma}W^{\delta I}.
\end{eqnarray}
\end{subequations}Note that the last line of \eqref{newvariationH}
vanishes automatically after the identification in \eqref{eq:F=00003DWH}.

As a final consistency check, it is not difficult to show that $T_{+}$
satisfies
\begin{equation}
\{T_{+}(\sigma'),T_{+}(\sigma)\}_{P.B.}=2T_{+}\,\partial_{\sigma}\delta(\sigma'-\sigma)+\partial_{\sigma}T_{+}\,\delta(\sigma'-\sigma),
\end{equation}
which demonstrates the last necessary condition for the nilpotency
of the BRST charge at the classical level.

\section{Discussion\label{sec:discussion}}

\

It is possible to show that the constraints displayed in \eqref{eq:nilpotencySUGRAconstraints},
\eqref{eq:holoSUGRAconstraints}, \eqref{eq:nilpotencySYM} and \eqref{eq:holoSYM}
imply the ten-dimensional supergravity and super Yang-Mills equations
of motion. Instead of presenting these results, which for the pure
spinor superstring were originally obtained and detailedly studied
in \cite{Berkovits:2001ue}, we will discuss the particularities of
the infinite tension string model. 

As presented in subsection \ref{sub:Classical-analysis}, there are
in principle three independent conditions to check in order to ensure
classical nilpotency of the BRST charge,\begin{subequations}\label{eq:nilpotencyconditions}
\begin{eqnarray}
Q_{\lambda}^{2} & = & 0,\\
\{\lambda^{\alpha}d_{\alpha}(\sigma'),T_{+}(\sigma)\}_{P.B.} & = & 0,\\
\{T_{+}(\sigma'),T_{+}(\sigma)\}_{P.B.} & = & 2T_{+}\,\partial_{\sigma}\delta(\sigma'-\sigma)+\partial_{\sigma}T_{+}\,\delta(\sigma'-\sigma).
\end{eqnarray}
\end{subequations}The first one is identical to the condition on
the left-moving BRST charge of the usual pure spinor superstring and
not surprisingly provides the so-called nilpotency constraints, given
by
\begin{equation}
\lambda^{\alpha}\lambda^{\beta}T_{\alpha\beta}^{\hphantom{\alpha\beta}A}=\lambda^{\alpha}\lambda^{\beta}H_{A\alpha\beta}=\lambda^{\alpha}\lambda^{\beta}\lambda^{\gamma}R_{\alpha\beta\gamma}^{\hphantom{\alpha\beta\gamma}\delta}=\lambda^{\alpha}\lambda^{\beta}F_{\alpha\beta}^{I}=0,
\end{equation}
exactly as in \cite{Berkovits:2001ue}. The second condition can be
stated as
\begin{equation}
[Q_{\lambda},T_{+}]=0,\label{eq:QlambdaT+discussion}
\end{equation}
and leads to the constraints of \eqref{eq:holoSUGRAconstraints} and
\eqref{eq:holoSYM}, which were obtained in \cite{Berkovits:2001ue}
by requiring holomorphicity of the BRST current. In the present model,
holomorphicity plays no role when it comes to imposing constraints.
This is so because the heterotic background coupled action, given
by
\begin{equation}
S=\frac{1}{2\pi}\int d^{2}z\{\mathcal{P}_{a}\bar{\Pi}^{a}+d_{\alpha}\bar{\Pi}^{\alpha}-\Pi^{A}\bar{\Pi}^{B}B_{BA}+\bar{A}^{I}J_{I}+w_{\alpha}\bar{\nabla}\lambda^{\alpha}+\bar{b}\bar{\partial}\bar{c}\}+S_{C},
\end{equation}
is still chiral. Therefore the remaining constraints should manifest
themselves through the condition \eqref{eq:QlambdaT+discussion}.
To interpret it, it might be useful to recall that conformal symmetry
is preserved at the classical level, such that $[Q_{\lambda},T]=0$.
We are then left with
\begin{eqnarray}
[Q_{\lambda},\mathcal{H}] & = & 0,
\end{eqnarray}
\emph{cf}. the definition \ref{eq:T+-}. This is precisely the \emph{ad
hoc} condition used in \cite{Chandia-Vallilo} for the type II construction.
Here, however, it is naturally embedded in the BRST operator. From
the sectorized point of view, the condition above is equivalent to
the conservation of the BRST charge separately in each sector.

Concerning the third condition in \eqref{eq:nilpotencyconditions},
it was verified to hold \emph{independently} of the background constraints.
This is partially connected to the classical conformal symmetry but
we do not have a clear understanding so far. It implies, for example,
that
\begin{equation}
\{\mathcal{H}(\sigma'),\mathcal{H}(\sigma)\}_{P.B.}=2T\,\partial_{\sigma}\delta(\sigma'-\sigma)+\partial_{\sigma}T\,\delta(\sigma'-\sigma).
\end{equation}
We expect this relation to hold in the type II case as well.

An interesting observation is that the background can be absorbed
by a field redefinition in the action, such that
\begin{equation}
S=\frac{1}{2\pi}\int d^{2}z\{\bar{\partial}Z^{M}\boldsymbol{P}_{M}+w_{\alpha}\bar{\partial}\lambda^{\alpha}+\bar{b}\bar{\partial}\bar{c}\}+S_{C},
\end{equation}
where
\begin{eqnarray}
\boldsymbol{P}_{M} & \equiv & E_{M}^{\hphantom{M}a}(\mathcal{P}_{a}+\Pi^{A}B_{Aa}+A_{a}^{I}J_{I}+w_{\gamma}\lambda^{\beta}\Omega_{a\beta}^{\hphantom{a\beta}\gamma})\nonumber \\
 &  & -E_{M}^{\hphantom{M}\alpha}(d_{\alpha}-\Pi^{A}B_{A\alpha}-A_{\alpha}^{I}J_{I}-w_{\gamma}\lambda^{\beta}\Omega_{\alpha\beta}^{\hphantom{\alpha\beta}\gamma}).
\end{eqnarray}
In this case, one can work with $\boldsymbol{P}_{M}$ as a fundamental
field and rewrite the BRST charge by expressing $d_{\alpha}$ and
$\mathcal{P}_{a}$ as functions of $\boldsymbol{P}_{M}$ and the other
worldsheet fields. Therefore we have instead a free action and the
heterotic background appears as a deformation of the BRST charge,
supporting the observation made by Chandia and Vallilo that the vertex
operators in this model could be seen as flutuations of $\mathcal{H}$.
Needless to emphasize, quantum consistency of the theory would be
much easier to verify in this approach, similarly to what was done
in \cite{Adamo:2014wea} for the NS-NS background. As in there, we
expect the dilaton superfield to start playing a fundamental role
in the quantum formulation of the theory.

It should be noted that in the original ambitwistor strings, either
in RNS or with pure spinors, the heterotic supergravity sector has
some unsolved issues. For example, Mason and Skinner \cite{Mason:2013sva}
computed the n-particle tree level amplitude and could not interpret
them in terms of standard space-time gravity, with the 3-point amplitude
suggesting a $(\textrm{Weyl})^{3}$-type vertex. On the other hand,
the supergravity vertex of \cite{Jusinskas:2016qjd} in the sectorized
string seems to provide the correct OPE structure in the 3-point amplitude,
resembling the usual heterotic pure spinor string up to numerical
factors. This subject deserves a deeper investigation and might shed
some light on the model. Naturally, if we go to 4-point amplitudes
or higher we need also the integrated vertices. We still do not have
a simple proposal for such operators. However, there are interesting
hints pointing out that the holomorphic sectorization can be extended
the bosonic string and to the Ramond-Neveu-Schwarz and Green-Schwarz
formalisms \cite{Chandia-Jusinskas}. We hope that understanding this
construction will provide a better basis to approach the problem of
the integrated vertex operator in the infinite tension limit using
pure spinors. Once this step is taken, we will finally be able to
compute the tree level amplitudes to compare them with the Cachazo-He-Yuan
formulae \cite{Cachazo:2013hca} and investigate the modular invariance
of the theory at 1-loop, for example, as done in \cite{Adamo:2013tsa}.

\

\textbf{Acknowledgements:} TA acknowledges financial support from
Conselho Nacional de Desenvolvimento Cient\'ifico e Tecnol\'ogico (CNPq).
The research of TA was also supported in part by the Knut and Alice
Wallenberg Foundation under grant 2015.0083. RLJ would like to thank
the Grant Agency of the Czech Republic for financial support under
the grant P201/12/G028. 

\appendix

\section{Superspace notation and conventions\label{sec:notation}}

In this work, we use the following sets of indices:

\begin{center}

\begin{tabular}{rl}
$a,b,\ldots=0$ to 9 :  & ten-dimensional tangent space vector, \tabularnewline
$\alpha,\beta,\ldots=1$ to 16 :  & ten-dimensional tangent space chiral spinor, \tabularnewline
$m,n,\ldots=0$ to 9 :  & ten-dimensional manifold vector, \tabularnewline
$\mu,\nu,\ldots=1$ to 16 :  & ten-dimensional manifold chiral spinor, \tabularnewline
$A,B,\ldots$  & collectively denote $(a,\alpha),(b,\beta),\ldots\,,$ \tabularnewline
$M,N,\ldots$  & collectively denote $(m,\mu),(n,\nu),\ldots\,.$ \tabularnewline
\end{tabular}

\end{center}

Our conventions concerning differential forms are the same as those
in \cite{Wess:1992cp}. In particular, given a manifold with vielbein
$E$ and connection $\Omega$, we define the torsion to be 
\begin{equation}
T=\mathrm{d}E+E\wedge\Omega,
\end{equation}
or, in components,
\begin{equation}
T_{NM}^{\hphantom{NM}A}=2\partial_{[N}E_{M)}^{\hphantom{M)}A}+(-1)^{N(B+M)}E_{M}^{\hphantom{M}B}\Omega_{NB}^{\hphantom{NB}A}-(-1)^{MB}E_{N}^{\hphantom{N}B}\Omega_{MB}^{\hphantom{MB}A},
\end{equation}
where the graded symmetrization is defined as
\begin{equation}
\Xi_{[NM)}\equiv\frac{1}{2}\Big[\Xi_{NM}-(-1)^{NM}\Xi_{MN}\Big],
\end{equation}
and the indices appearing at the exponents should be replaced by their
gradings, \emph{i.e.} $+1$ for spinorial indices and $0$ otherwise.
As usual, one can write the torsion in terms of tangent space indices
by contracting it with vielbeins: 
\begin{equation}
T_{BC}^{\hphantom{BC}A}=(-1)^{B(C+M)}E_{C}^{\hphantom{C}M}E_{B}^{\hphantom{B}N}T_{NM}^{\hphantom{NM}A}.
\end{equation}

Another important quantity is the curvature tensor, defined in terms
of the connection as 
\begin{equation}
R=\mathrm{d}\Omega+\Omega\wedge\Omega,
\end{equation}
or, in components, 
\begin{equation}
R_{NMA}^{\hphantom{NMA}B}=2\partial_{[N}\Omega_{M)A}^{\hphantom{M)A}B}+(-1)^{N(A+C+M)}\Omega_{MA}^{\hphantom{MA}C}\Omega_{NC}^{\hphantom{NC}B}-(-1)^{M(A+C)}\Omega_{NA}^{\hphantom{NA}C}\Omega_{MC}^{\hphantom{MC}B}.
\end{equation}

Finally, the 3-form $H=\mathrm{d}B$ is given in components by $H_{MNP}=3\partial_{[M}B_{NP)}$.

\section{$SO(32)$ realization of the gauge sector\label{sec:currents}}

Here we will present a realization of the action $S_{C}$ describing
the gauge sector of the heterotic string, focusing on the $SO(32)$
group, which has a simpler construction.

Concerning notation, the vector and the adjoint representations of
the group, will be respectively denoted by the indices $i,j,k,\ldots=1,\ldots,32$
and $I,J,K,\ldots=1,\ldots,496$. The metric set ($\delta^{ij}$,
$\delta^{IJ}$, $\delta_{ij}$, $\delta_{IJ}$) will be used to raise
and to lower the group indices.

The generators of the $SO(32)$ group will be denoted by the anti-Hermitian
operators $T^{I}$. The algebra can be cast as
\begin{equation}
[T_{I},T_{J}]=f_{IJ}^{\hphantom{IJ}K}T_{K},
\end{equation}
where $f_{IJK}=f_{IJ}^{\hphantom{IJ}L}\delta_{LK}$ are real and totally
antisymmetric structure constants constrained to satisfy the Jacobi
identity
\begin{equation}
f_{IJ}^{\hphantom{IJ}M}f_{MK}^{\hphantom{MK}L}+f_{JK}^{\hphantom{JK}M}f_{MI}^{\hphantom{MI}L}+f_{KI}^{\hphantom{KI}M}f_{MJ}^{\hphantom{MJ}L}=0.
\end{equation}

The action $S_{C}$ consists of a (free) set of $32$ real worldsheet
fermions, $\psi_{i}$, such that
\begin{equation}
S_{C}=\frac{1}{4\pi}\int d^{2}z\,\psi^{i}\bar{\partial}\psi^{j}\delta_{ij}.
\end{equation}
The associated energy-momentum tensor is
\begin{equation}
T_{C}=-\frac{1}{2}\psi^{i}\partial\psi^{j}\delta_{ij}.
\end{equation}
Using the simple OPE
\begin{equation}
\psi^{i}(z)\psi^{j}(y)\sim\frac{\delta^{ij}}{(z-y)},
\end{equation}
one can easily compute
\begin{equation}
T_{C}(z)T_{C}(y)\sim\frac{(\frac{16}{2})}{(z-y)^{4}}+\frac{2T_{C}}{(z-y)^{2}}+\frac{\partial T_{C}}{(z-y)},
\end{equation}
showing that the central charge of the system is $16$, as required
by the vanishing conformal anomaly in the heterotic string.

The $SO(32)$ group structure in the worldsheet theory can be seen
through the current
\begin{equation}
J_{I}\equiv\frac{1}{2}(T_{I}^{jk}\psi_{j}\psi_{k}).
\end{equation}
Observe that $J^{I}$ is conserved, $\bar{\partial}J^{I}=0$,
which follows from the classical equation of motion for $\psi_{i}$.
For completeness, we can mention that the currents $J_{I}$ define
an Affine Lie algebra at the quantum level, which can be read from
the OPE:
\begin{equation}
J_{I}(z)J_{J}(y)\sim\frac{1}{2}\frac{\textrm{Tr}(T_{I}T_{J})}{(z-y)^{2}}+f_{IJ}^{\hphantom{IJ}K}\frac{J_{K}}{(z-y)}.
\end{equation}

Following the notation of subsection \ref{sub:Classical-analysis},
we can see that the canonical conjugate of $\psi^{i}$ is identified
with $\psi^{i}$ itself, as usual for fermionic systems. Therefore
we have to use Dirac's procedure to deal with this constraint in order
to obtain the Dirac brackets for $\psi^{i}$, given by
\begin{equation}
\{\psi^{i}(\sigma'),\psi^{j}(\sigma)\}=\delta^{ij}\delta(\sigma-\sigma'),
\end{equation}
and show that the currents satisfy
\begin{eqnarray}
\{T_{C}(\sigma'),J_{I}(\sigma)\} & = & J_{I}\partial_{\sigma}\delta(\sigma-\sigma')+\partial_{\sigma}J_{I}\delta(\sigma'-\sigma),\\
\{J_{I}(\sigma'),J_{J}(\sigma)\} & = & f_{IJ}^{\hphantom{IJ}K}J_{K}\delta(\sigma'-\sigma),
\end{eqnarray}
which are used in section \ref{sec:background}.

The simplest interacting model involving the currents $J_{I}$ is
the minimal coupling to an external source $\bar{A}_{I}$, such
that
\begin{equation}
S_{C}^{int}=S_{C}+\frac{1}{2\pi}\int d^{2}z\,J_{I}\bar{A}^{I}.\label{eq:Scint}
\end{equation}
In this case, the equation of motion for the current can be determined
to be
\begin{equation}
\bar{\partial}J_{I}+f_{IJ}^{\hphantom{IJ}K}\bar{A}^{J}J_{K}=0.\label{eq:eomJi}
\end{equation}
It is interesting to observe that this coupling has a very natural
symmetry at the classical level. Consider the transformation
\begin{equation}
\delta\bar{A}^{I}=\bar{\partial}\Sigma^{I}-f_{JK}^{\hphantom{JK}I}\Sigma^{J}\bar{A}^{K},
\end{equation}
where $\Sigma^{I}$ is a generic parameter in the adjoint representation
of $SO(32)$. It is straightforward to show that $S_{C}^{int}$ transforms
as
\begin{equation}
\delta S_{C}^{int}=-\frac{1}{2\pi}\int d^{2}z(\bar{\partial}J_{I}+f_{IJ}^{\hphantom{IJ}K}\bar{A}^{J}J_{K})\Sigma^{I},
\end{equation}
which vanishes for classical configurations of $J_{I}$, \emph{cf}.
equation \eqref{eq:eomJi}.

\end{document}